\documentclass{PoS}
\usepackage{graphicx}
\usepackage{dcolumn}
\usepackage{bm}
\usepackage{amssymb}
\usepackage{amsmath}
\usepackage{epsfig}    
\usepackage{color}
\usepackage{slashed}
\title{Light charged Higgs boson scenario in 3HDMs}

\ShortTitle{Charged Higgs Searches@LHC}

\author{A.~G.~Akeroyd\\
        School of Physics and Astronomy, University of Southampton,\\
	Southampton, SO17 1BJ, United Kingdom\\
        E-mail: \email{a.g.akeroyd@soton.ac.uk}}

\author{Stefano Moretti\\
        School of Physics and Astronomy, University of Southampton,\\
	Southampton, SO17 1BJ, United Kingdom\\
        E-mail: \email{s.moretti@soton.ac.uk}}

\author{\speaker{Kei Yagyu}\\
INFN, Sezione di Firenze, and Department of Physics and Astronomy, University of Florence, Via
G. Sansone 1, 50019 Sesto Fiorentino, Italy\\
        E-mail: \email{yagyu@fi.infn.it}}

\author{Emine Yildirim\\
        School of Physics and Astronomy, University of Southampton,\\
	Southampton, SO17 1BJ, United Kingdom\\
        E-mail: \email{ey1g13@soton.ac.uk}}

\abstract{
The measurement of the $B\to X_s\gamma$ process gives 
important constraints on physics related to charged Higgs bosons ($H^\pm$).
In 2-Higgs Doublet Models (2HDMs) with a softly-broken $Z_2$ symmetry, a light $H^\pm$ scenario, 
in which $H^\pm$ can be produced via the top decay, is possible in two of four types of Yukawa interactions (the so-called Type-I and Type-X). 
In these types of 2HDMs, the $H^\pm \to \tau^\pm \nu$ decay mode is dominant in wide regions of the parameter space. 
In this report, we discuss the other possibility of a light charged Higgs boson scenario in 3-Higgs-Doublet Models (3HDMs) based on the results obtained in Ref.~{\cite{3hdm}}. 
We show that charged Higgs bosons can mainly decay into $cb$ without contradiction with 
the $B\to X_s\gamma$ data and the direct searches for charged Higgs bosons at the LHC, and this scenario cannot be realized in the 2HDMs.
}

\FullConference{Prospects for Charged Higgs Discovery at Colliders\\
		3-6 October 2016\\
		Uppsala, Sweden}

\begin{document}

\section{Introduction}

After the LHC Run-I experiments, a Higgs boson with a mass about 125 GeV has been discovered in various channels such as $\gamma\gamma$, $ZZ^*$, $WW^*$ and $\tau^+\tau^-$ modes whose 
event rates are consistent with those predicted by the Standard Model (SM)~\cite{LHC}. 
From this experimental results, there is almost no doubt about the existence of at least one isospin doublet scalar field.  

Now, the natural question is whether the true Higgs sector is composed of a single doublet or multi doublets. 
There are several reasons which support the possibility of the multi doublet case. 
Firstly, multi doublet models predict the electroweak $\rho$ parameter to be unity at the tree level without any parameter tuning, which 
is quite consistent with its experimental value. 
Secondly, these models can reproduce various predictions in the SM by taking an appropriate limit, so that 
the current measurements of the Higgs boson properties at the LHC can be explained. 
Finally, when we consider new physics models beyond the SM, the multi doublet structure often appears in their Higgs sector. 
As the well known example, the Minimal Supersymmetric SM (MSSM) requires two doublets in its Higgs sector for the gauge anomaly cancellation. 
Therefore, as a bottom-up approach 
it is important to study properties of multi doublet models not only to 
elucidate the Higgs sector but also to extract information on new physics models. 

One of the characteristic features of multi doublet models is the appearance of physical charged Higgs bosons, so that 
their detection is quite important to test such scenarios. 
Properties of charged Higgs bosons strongly depend on the construction of the model, e.g., the structure of Yukawa interactions. 
In this report, we focus on the phenomenology of charged Higgs bosons in a model with Natural Flavour Conservation (NFC)~\cite{NFC}, where 
each mass of three types of fermions, i.e., up-type quarks, down-type quarks and charged leptons is given by only one of the doublets. 
This scenario can be naturally realized by imposing a discrete symmetry in the Higgs sector. 
In particular, we clarify the difference between the nature of charged Higgs bosons in 2HDMs and 3HDMs. 

This report is organized as follows. 
In Sec.~II, we briefly review the charged Higgs boson sector in 2HDMs and 3HDMs.
In Sec.~III, we discuss constraints on the parameter space from $B\to X_s\gamma$ and the direct searches at the LHC.  
Conclusions and discussions are given in Sec.~IV. 

\section{Model}

We consider 2HDMs and 3HDMs, in which the Higgs sector is composed of two isospin doublets $\Phi_1$ and $\Phi_2$
and three doublets $\Phi_1$, $\Phi_2$ and $\Phi_3$, respectively. 
We assume CP-conservation of the Higgs sector for simplicity. 
In these models, the sum rule for the Vacuum Expectation Values (VEVs) is satisfied as $v^2\equiv \sum_a v_a^2 = (\sqrt{2}G_F)^{-1/2}$ ($a=1,2$ for 2HDMs and $a=1$--3 
for 3HDMs), where $v_a = \sqrt{2}\langle \Phi_a \rangle$ and $G_F$ is the Fermi constant. 
It is convenient to introduce the ratio of the VEVs as follows
\begin{align}
&\tan\beta \equiv \frac{v_2}{v_1}~~\text{for~2HDMs},\quad
\tan\beta \equiv \frac{v_2}{\sqrt{v_1^2+v_3^2}},~\tan\gamma \equiv \frac{v_3}{v_1}~\text{for~3HDMs}. 
\end{align}

In the scenario based on NFC, the Yukawa Lagrangian is given by the following form:
\begin{align}
-{\cal L}_Y = Y_u \bar{Q}_L (i\sigma_2)\Phi_u^* u_R^{} 
+Y_d \bar{Q}_L \Phi_d d_R^{}
+Y_e \bar{L}_L \Phi_e e_R^{} + \text{h.c.}, \label{yuk}
\end{align}
where $\Phi_{u}$, $\Phi_{d}$ and $\Phi_{e}$ are either $\Phi_1$ or $\Phi_2$ ($\Phi_1$, $\Phi_2$ or $\Phi_3$) in 2HDMs (3HDMs). 
This Lagrangian can be naturally realized by imposing a discrete symmetry, e.g., $Z_2$ and $Z_2\times Z_2$ in 2HDM and in 3HDMs, respectively, where 
these can be softly-broken by dimensionful scalar couplings in the scalar potential.  
In the 2HDM (3HDM), there are four (five) independent ways to construct the above Lagrangian depending on the choice of $\Phi_{u,d,e}$. 
In Table~\ref{Tab:type}, we define  four (five) types of Yukawa interactions in the 2HDM (3HDM), where Type-Z is allowed only in the 3HDM. 
A similar classification has also been done in Refs.~\cite{Grossman,Silva}. 

In the following, we first give the expressions of 
the interaction terms among charged Higgs bosons and fermions in 3HDMs and then 
we explain how those of the 2HDMs can be obtained. 
The interaction terms are extracted from Eq.~(\ref{yuk}) as follows:
\begin{align}
&-{\cal L}_Y^{\text{int}} = \frac{\sqrt{2}}{v}\sum_{a=1,2}\left(\bar{u}^j V_{ji} m_{d^i} X_a P_R  d^i  
+\bar{u}^im_{u^i} V_{ij} Y_a P_L  d^j 
+\bar{\nu}^i  m_{e^i} Z_a P_R  e^i \right)H_a^+   + \text{h.c.}, \label{yuk2a}
\end{align}
where $H_a^\pm$ ($a=1,2$) are physical charged Higgs bosons and $V_{ij}$ are the Cabibbo-Kobayashi-Maskawa matrix elements. 
The coefficients $X_a$, $Y_a$ and $Z_a$ are given by 
\begin{align}
X_1 &= \xi_1^d c_C^{} + \xi_2^ds_C^{},~ 
Y_1 = -(\xi_1^u c_C^{} + \xi_2^u s_C^{}),~
Z_1 = \xi_1^ec_C^{} + \xi_2^e s_C^{},  \label{xyz1}\\
X_2 &= -\xi_1^d s_C^{} + \xi_2^dc_C^{},~ 
Y_2 = -(-\xi_1^u s_C^{} +\xi_2^uc_C^{}),~
Z_2 = -\xi_1^es_C^{} + \xi_2^ec_C^{},  \label{xyz2}
\end{align}
where $s_C^{}=\sin\theta_C$ and $c_C^{}=\cos\theta_C$ with $\theta_C$ being the mixing angle between $H_1^\pm$ and $H_2^\pm$. 
In Eqs.~(\ref{xyz1}) and (\ref{xyz2}), the factors $\xi_a^f$ ($f=u,d,e$) depend on the type of Yukawa interactions, of which explicit forms
are given in Table~\ref{Tab:type} in terms of $\beta$ and $\gamma$. 
In 2HDMs, there is only one pair of charged Higgs bosons $H^\pm$, so that the coefficients $X_a$, $Y_a$ and $Z_a$ can simply be written as 
$X$, $Y$ and $Z$, respectively, and their expressions are obtained from $X_1$, $Y_1$ and $Z_1$ by taking $\theta_C \to 0$. 

\begin{table}[t]
\begin{center}
\begin{tabular}{c||c|c|c||ccc||ccc}
\hline\hline 
& $\Phi_u$ &$\Phi_d$ &$\Phi_e$ & $\xi_1^u$ & $\xi_1^d$  & $\xi_1^e$ & $\xi_2^u$& $\xi_2^d$& $\xi_2^e$\\\hline
Type-I   & $\Phi_2$ & $\Phi_2$ & $\Phi_2$ &$\cot\beta$&$\cot\beta$&$\cot\beta$&0&0&0  \\   
Type-II  & $\Phi_2$ & $\Phi_1$ & $\Phi_1$ &$\cot\beta$&$-\tan\beta$&$-\tan\beta$&0&$-\tan\gamma/\cos\beta$&$-\tan\gamma/\cos\beta$  \\  
Type-X   & $\Phi_2$ & $\Phi_2$ & $\Phi_1$  &$\cot\beta$&$\cot\beta$ &$-\tan\beta$&0&0&$-\tan\gamma/\cos\beta$ \\
Type-Y   & $\Phi_2$ & $\Phi_1$ & $\Phi_2$ &$\cot\beta$&$-\tan\beta$ &$\cot\beta$&0&$-\tan\gamma/\cos\beta$&0  \\
Type-Z   & $\Phi_2$ & $\Phi_1$ & $\Phi_3$  &$\cot\beta$&$-\tan\beta$  &$-\tan\beta$&0&$-\tan\gamma/\cos\beta$&$\cot\gamma/\cos\beta$\\
\hline\hline
\end{tabular} 
\end{center}
\caption{Four (Five) independent choices of the combination of $\Phi_u$, $\Phi_d$ and $\Phi_e$ named as Type-I, -II, -X, -Y and -Z in the 2HDM (3HDM). 
The $\xi_a^f$ ($a=1,2$ and $f=u,d,e$) factors in Eqs.~(2.4) and (2.5) are also shown in all types of Yukawa interactions. } 
\label{Tab:type}
\end{table}

The relevant parameters for the charged Higgs bosons are $\tan\beta$ and the mass $m_{H^\pm}$ in 2HDMs, while 
these are $\tan\beta$, $\tan\gamma$, $m_{H_1^\pm}$ (the mass of $H_1^\pm$), $m_{H_2^\pm}$ (the mass of $H_2^\pm$) and $\theta_C$ in 3HDMs, where 
we define $m_{H_2^\pm} \geq m_{H_1^{\pm}}$. 
We note that, if we introduce CP-violating couplings in the Higgs potential, a CP-phase appears in the mass matrix for the charged Higgs bosons in the 3HDMs, which 
is taken to be zero as we already assumed CP-conservation of the Higgs sector. 

\begin{figure}[t]
\begin{center}
\includegraphics[width=38mm]{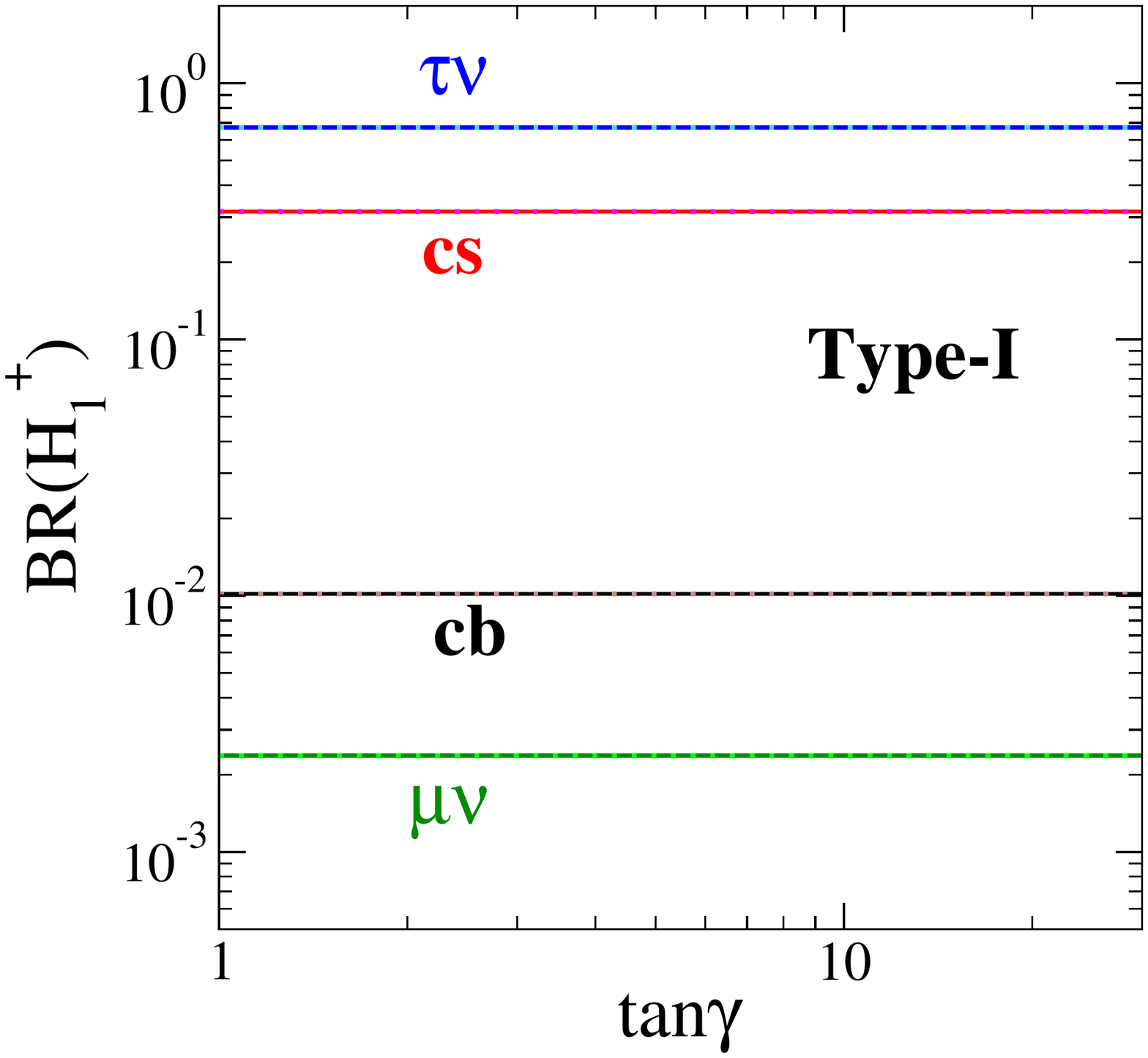}\hspace{-12mm}
\includegraphics[width=38mm]{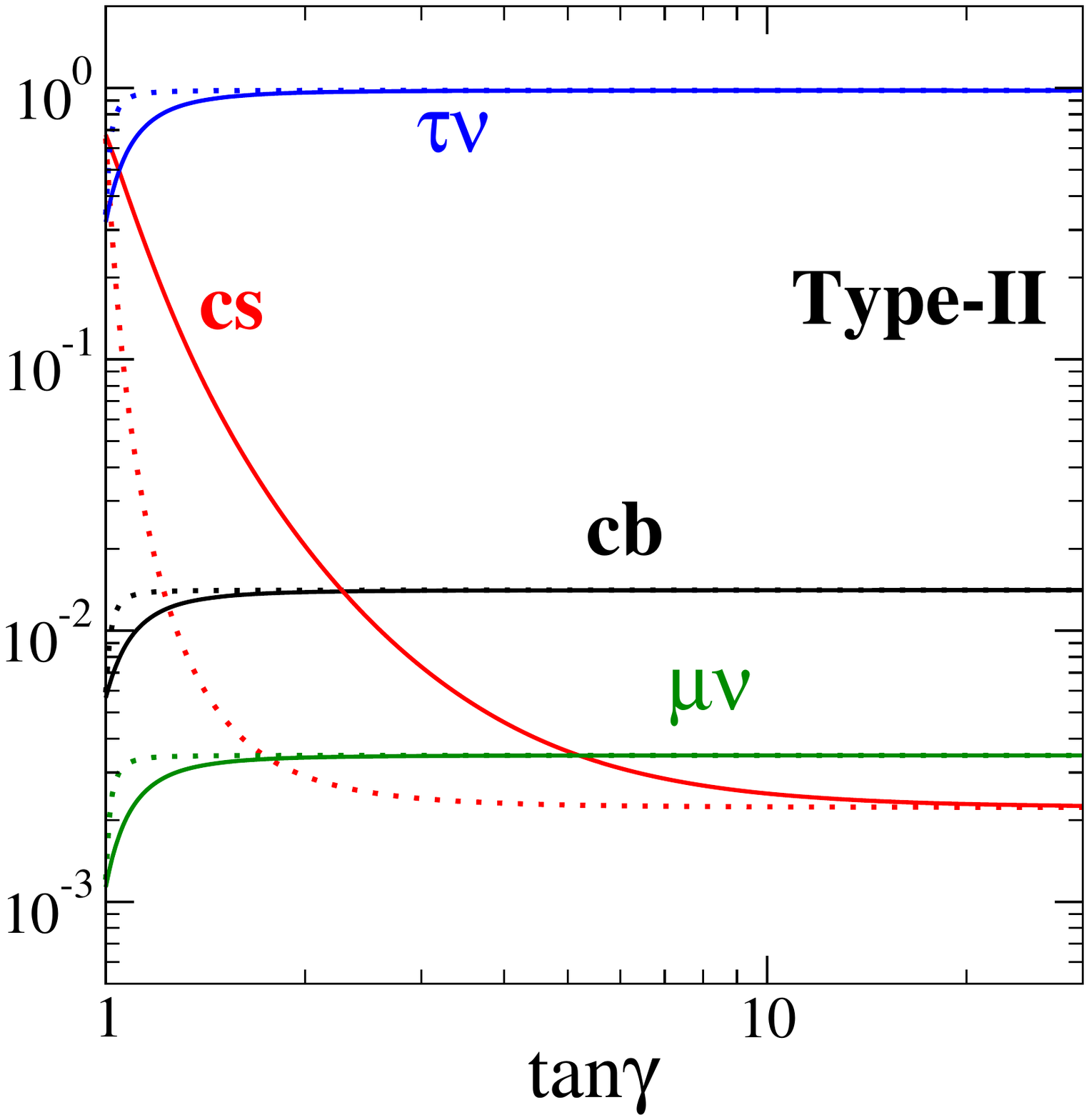}\hspace{-12mm}
\includegraphics[width=38mm]{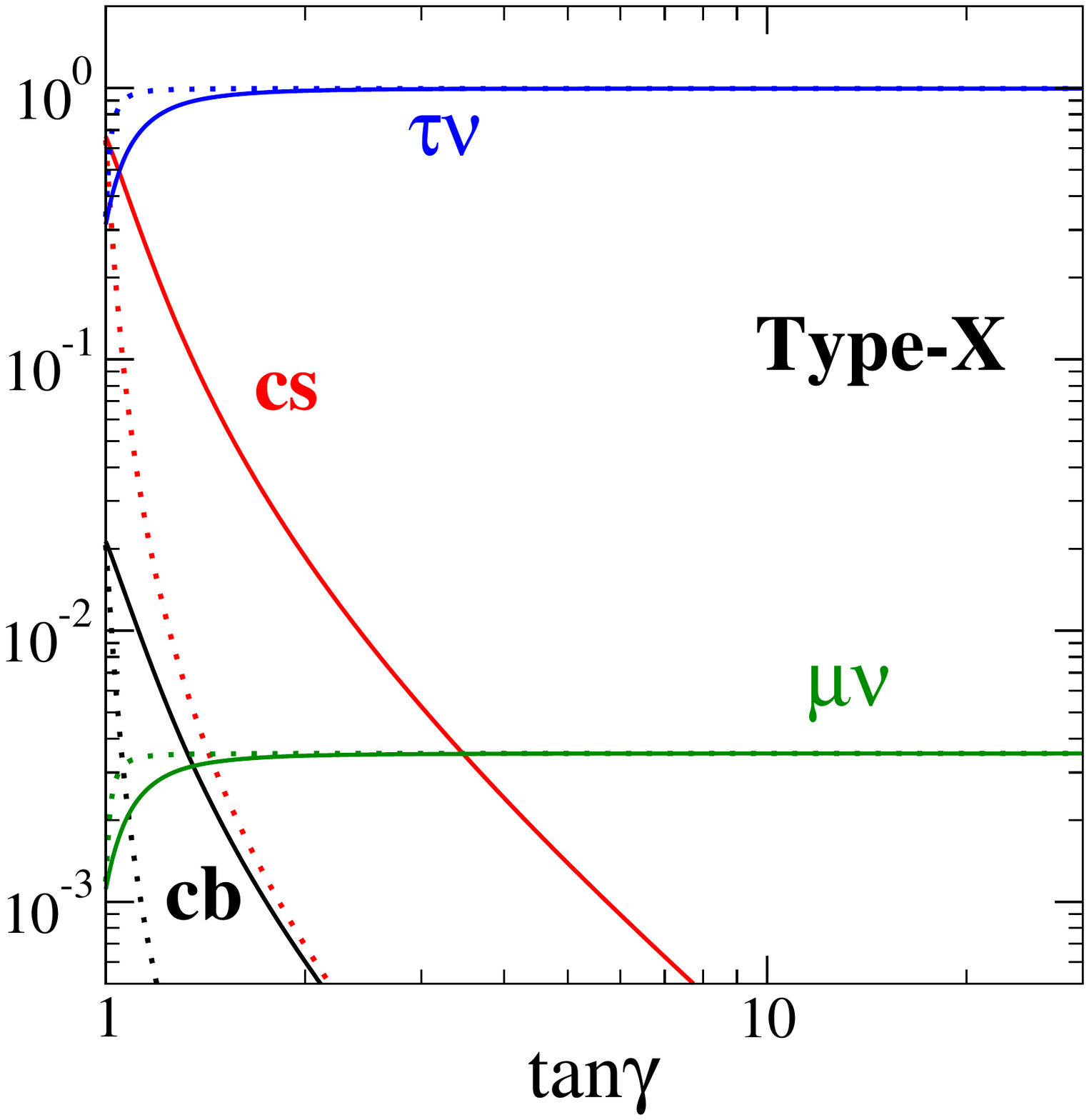}\hspace{-12mm}
\includegraphics[width=38mm]{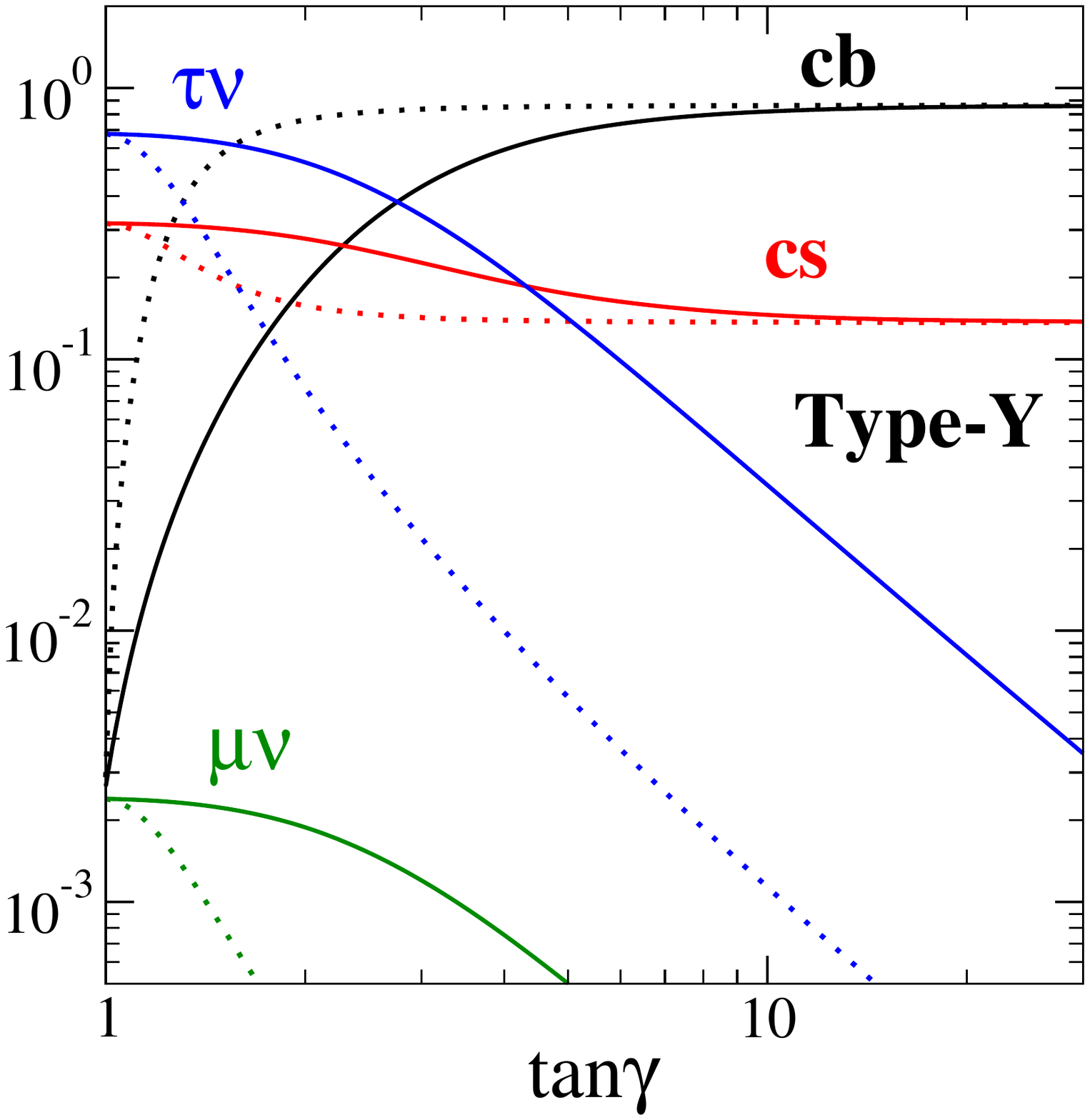}\hspace{-12mm}
\includegraphics[width=38mm]{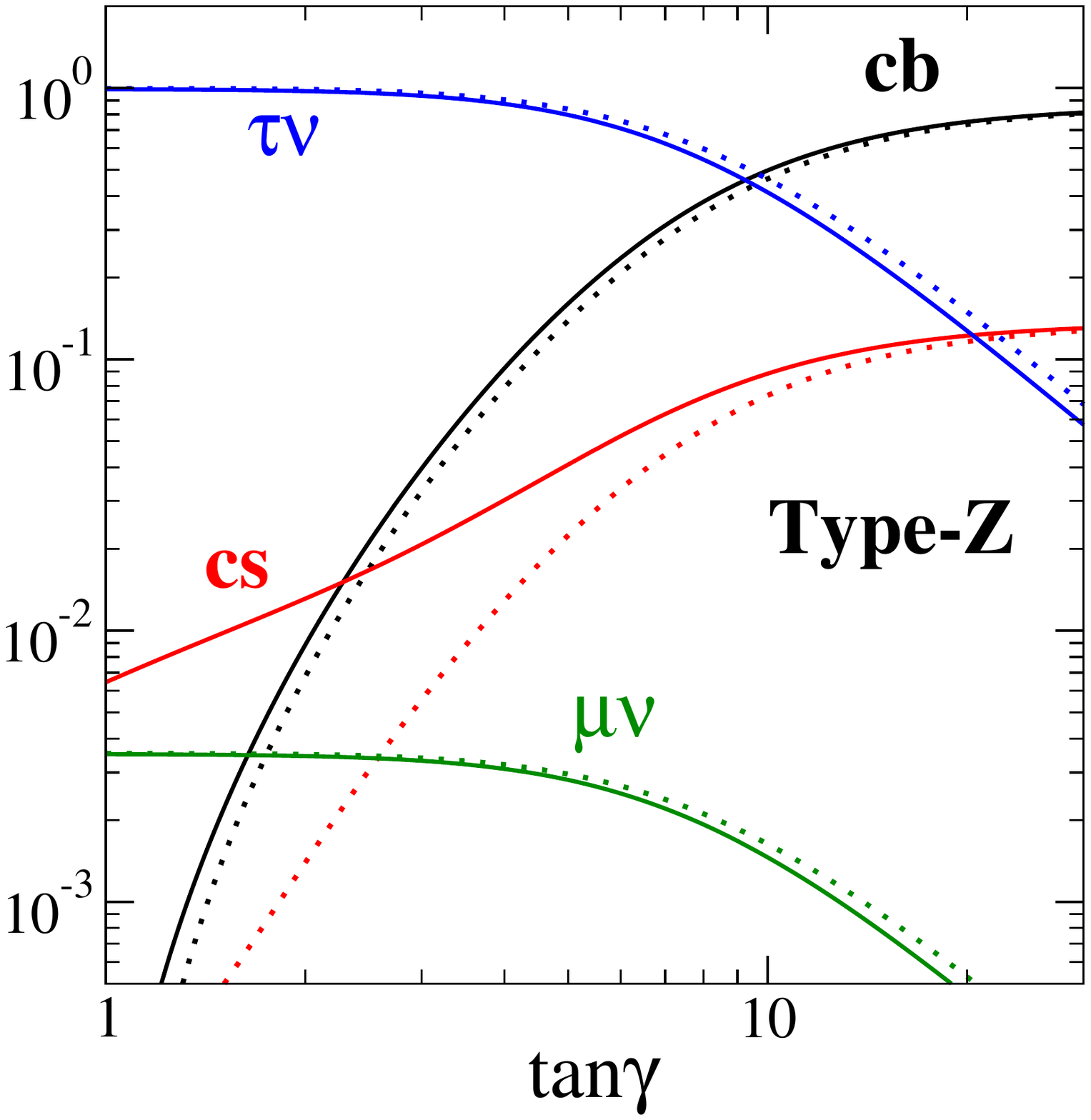}
\caption{BRs of $H_1^\pm$ as a function of $\tan\gamma$ in the Type-I, -II, -X, -Y and -Z 3HDM
from the left to right panels in the case for $m_{H_1^\pm}=150$ GeV and $\theta_C= -\pi/4$.  
We take $\tan\beta=2(5)$ for the solid (dotted) curves. 
}
\label{br}
\end{center}
\end{figure}

In Fig.~\ref{br}, we show the Branching Ratios (BRs) of the lighter charged Higgs bosons $H_1^\pm$ as a function of $\tan\gamma$ in all the five types of Yukawa interactions. 
We here assume all the other extra neutral Higgs bosons to be heavier than $H_1^\pm$, so that the decay modes into a neutral state are kinematically not allowed. 
In addition, we take the alignment limit, i.e., 
the SM-like Higgs boson $h$ exactly corresponds to the CP-even component scalar field in the doublet including the VEV $v$ in the Higgs basis. 
In this limit, the coupling $H_1^\pm h W_\mu^\mp$ vanishes at the tree level. 
We can see that the $H_1^\pm \to \tau^\pm \nu$ mode can be dominant in the Type-I, -II and -X 3HDMs, while 
the $H_1^\pm \to cb$ mode can be important when $\tan\gamma\gtrsim 3(10)$ in the Type-Y (Type-Z) 3HDM for $\tan\beta = 2$. 
This results do not change so much when we take the other values of $m_{H_1^\pm}$ and $\theta_C^{}$ as long as $m_{H_1^\pm}< m_t$. 
It is important to mention here that the $H^\pm \to cb$ mode can also be dominant in the Type-Y 2HDM when $m_{H^\pm}< m_t$ (see Ref.~\cite{typeX}). 
However, the light charged Higgs boson scenario is not possible in the Type-Y 2HDM  
due to constraints from the $B\to X_s\gamma$ data as we see it in the next section. 

\section{Constraints on the parameter space}

We first review the constraint on the parameter space of 2HDMs and 3HDMs 
from the measurement of the rare decay process: $B\to X_s\gamma$. 
The measured value of its BR is given~\cite{HFAG} as 
\begin{align}
&{\rm BR}(B \to X_s\gamma) = (3.43 \pm 0.22)\times 10^{-4}. 
\end{align}

In multi doublet models, there are charged Higgs boson loop contributions in addition to the SM $W$ boson loop. 
The effective Lagrangian relevant to the $b\to s\gamma$ process is obtained 
after integrating out the heavy degrees of freedom such as $W$, $t$ and charged Higgs bosons as:
\begin{align}
{\cal L}_{\text{eff}} = \frac{4G_F}{\sqrt{2}}V_{ts}^* V_{tb} C_7(\mu){\cal O}_7(\mu), 
\end{align}
where ${\cal O}_7$ is the dimension six Wilson coefficient. When we neglect the strange quark mass, it can be written as 
\begin{align} 
{\cal O}_7(\mu) = \frac{e}{16\pi^2}\bar{m}_b(\mu)(\bar{s}_L\sigma^{\mu\nu}b_R)F_{\mu\nu}, 
\end{align}
where $F^{\mu\nu}$ is the field strength tensor for the photon and $\bar{m}_b(\mu)$ is the running bottom quark mass in the $\overline{\text{MS}}$ scheme at the scale $\mu$. 
All information of new physics is contained in the Wilson coefficient $C_7$. In the 3HDM, it is expressed by
\begin{align}
C_7(\mu,m_{H_1^\pm},m_{H_2^\pm}) = C_{7,\text{SM}}(\mu)
+\sum_{a=1,2}\left[(X_aY_a^*)C_{7,XY}(\mu,m_{H_a^\pm})+|Y_a|^2C_{7,YY}(\mu,m_{H_a^\pm})\right].   \label{wc}
\end{align}
The full analytic expressions for $C_{7,\text{SM}}$, $C_{7,XY}$ and $C_{7,YY}$ are given, e.g., in Ref.~\cite{Borz} 
at Next-To-Leading Order (NLO) in QCD~\footnote{These coefficients can be calculated at the matching scale $\mu_W^{}$, 
where the effective low energy theory matches with the full theory, 
and then those at the scale of the bottom quark can be derived according to the renormalization group running.}. 
The important thing for Eq.~(\ref{wc}) is that the relative sign of the three functions is the same with each other at least 
in the NLO calculation~\cite{Borz}. 
Thus, if the sign of $X_aY_a$ is positive (negative), then the contribution from $C_{7,XY}^{\text{eff}}(\mu,m_{H_a^\pm})$
becomes constructive (destructive) the other two contributions. 
In the 2HDM, the product $XY$ is given to be $-\cot^2\beta$ and $(+1)$ in the Type-I and -X (Type-II and -Y). 
Therefore, the charged Higgs boson contribution is destructive (constructive) in the Type-I and -X (Type-II and -Y) 2HDMs. 
In Ref.~\cite{Misiak}, the BR of $B\to X_s\gamma$ has been calculated in 2HDMs at Next-To-Next-To-Leading Order (NNLO) in QCD, where 
the 95\% CL lower limit on $m_{H^\pm}$ has been presented to be about 480 GeV in the Type-II Yukawa interaction, while in the Type-I 2HDM 
${\cal O}(100)$ GeV of $m_{H^\pm}$ is allowed when $\tan\beta\gtrsim 2.5$~\cite{Misiak2}. 

\begin{figure}[t]
\begin{center}
\includegraphics[width=50mm]{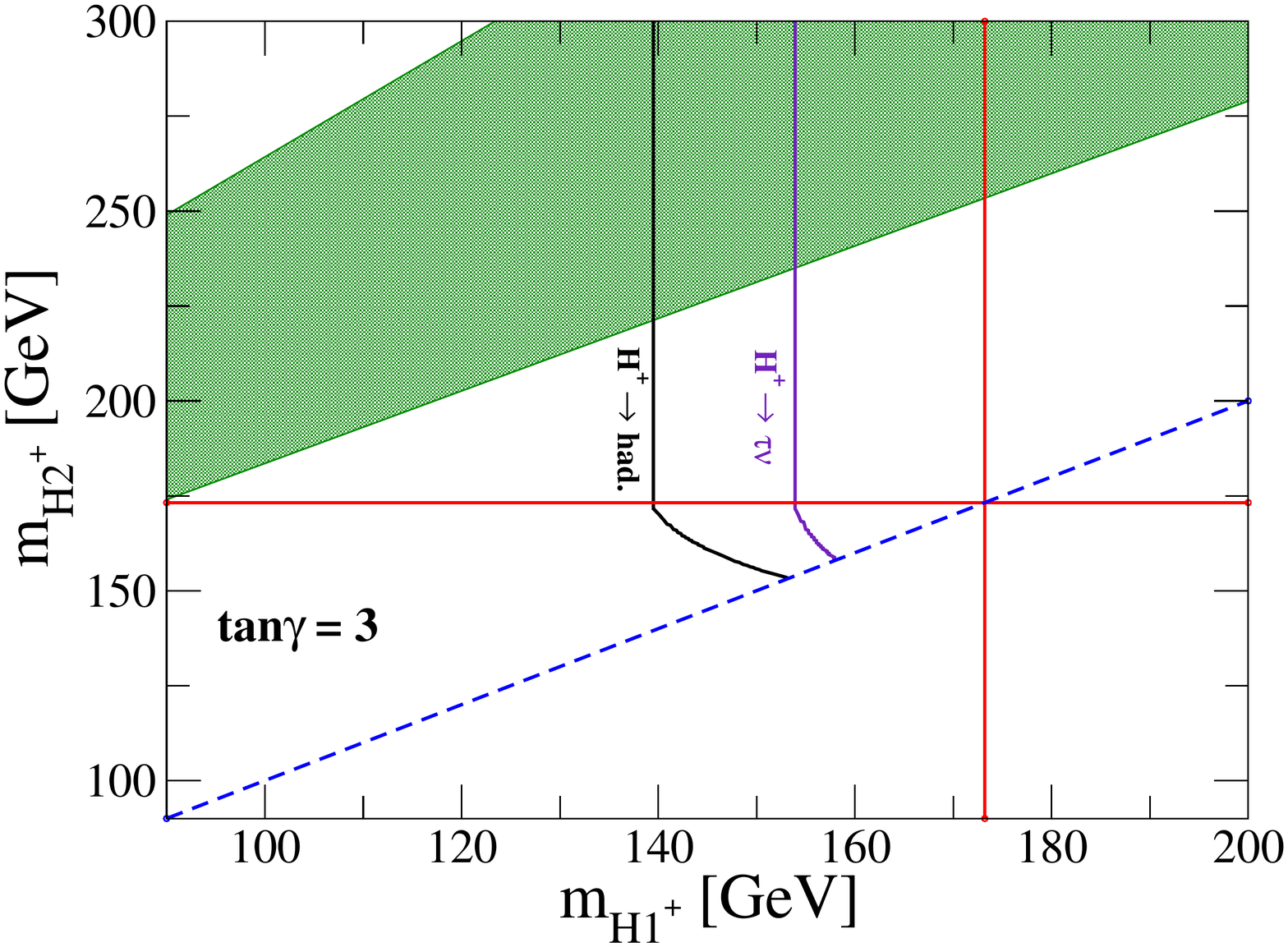}\hspace{-5mm}
\includegraphics[width=50mm]{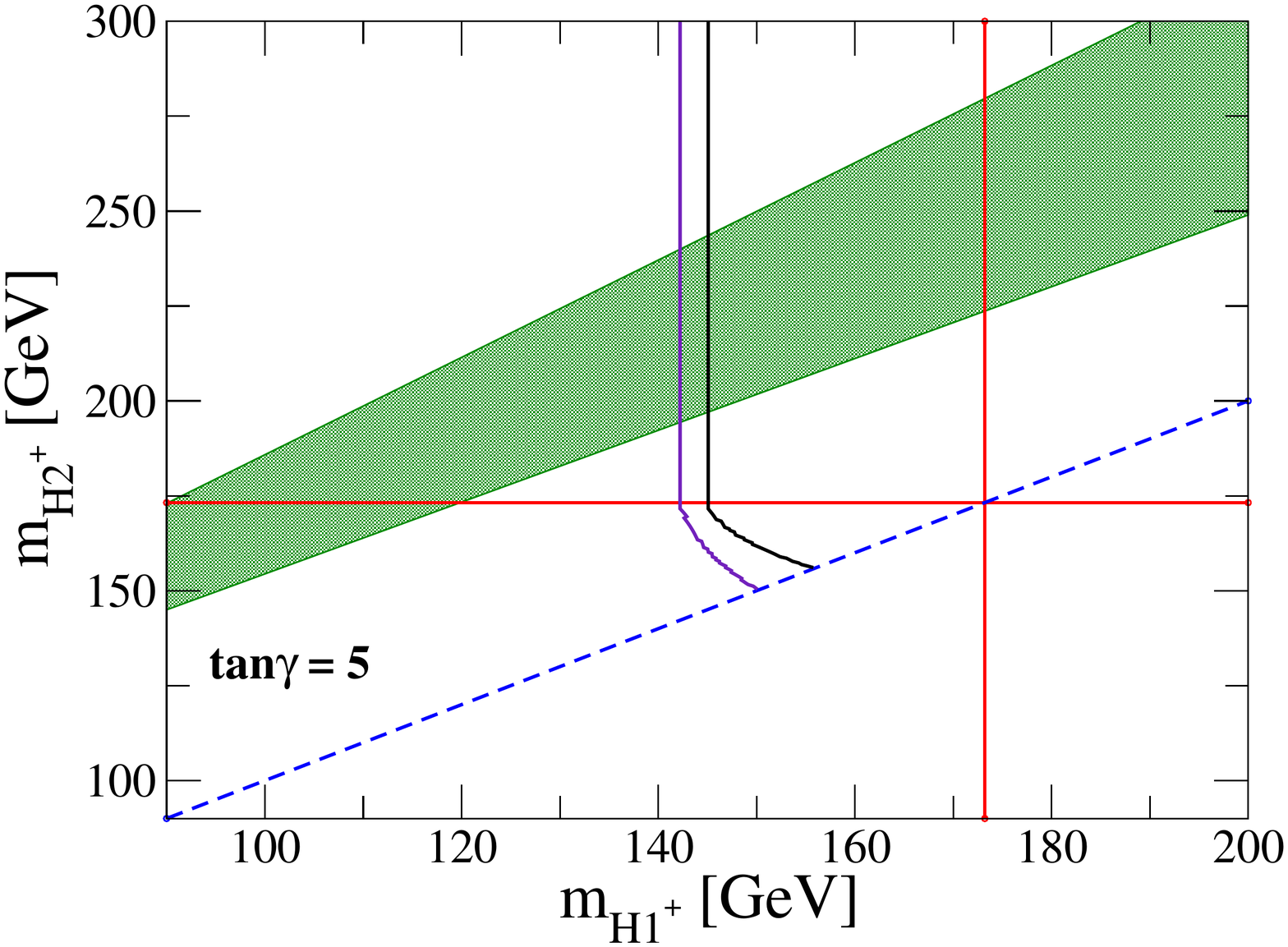}\hspace{-5mm}
\includegraphics[width=50mm]{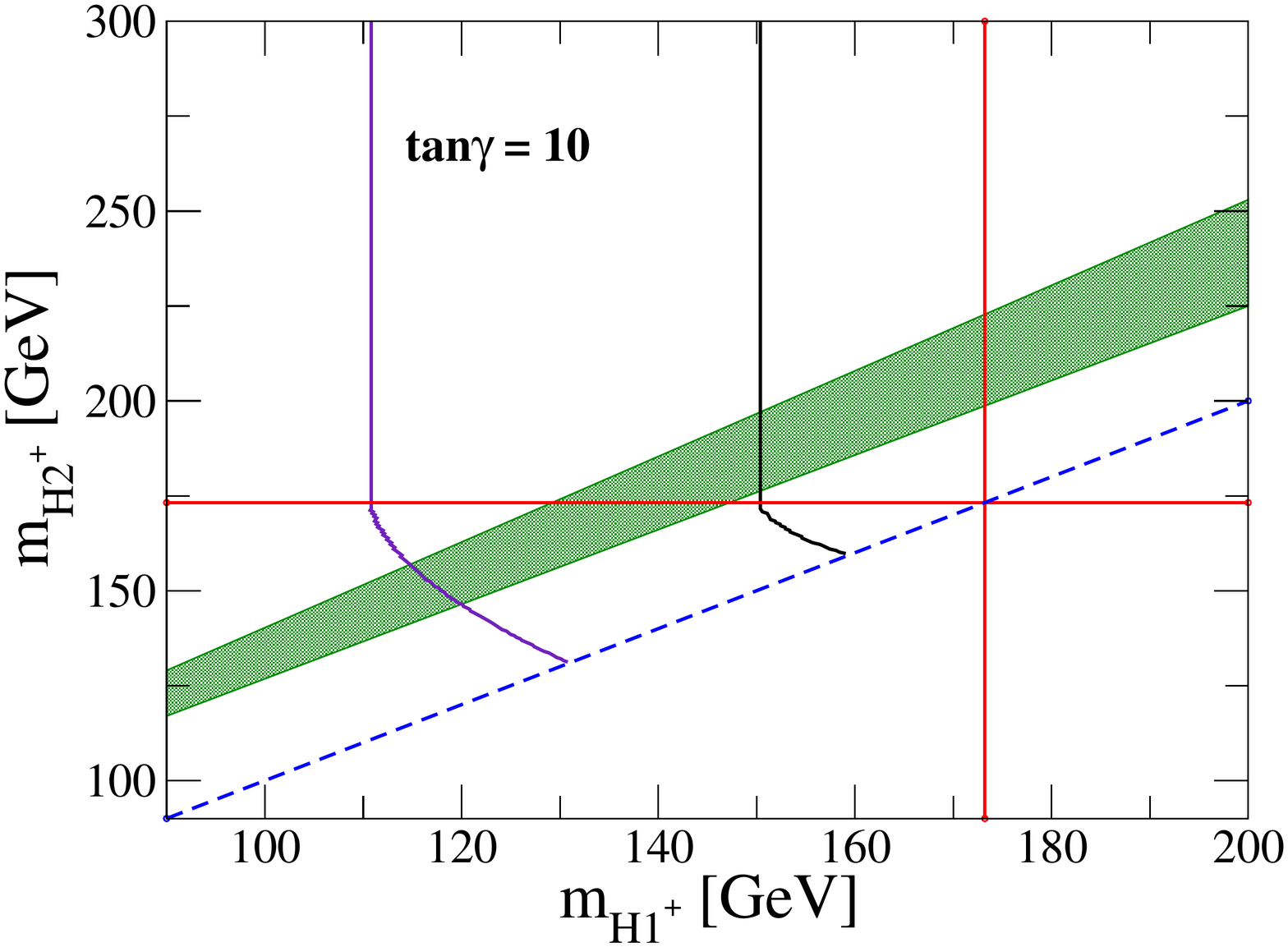} \\
\includegraphics[width=50mm]{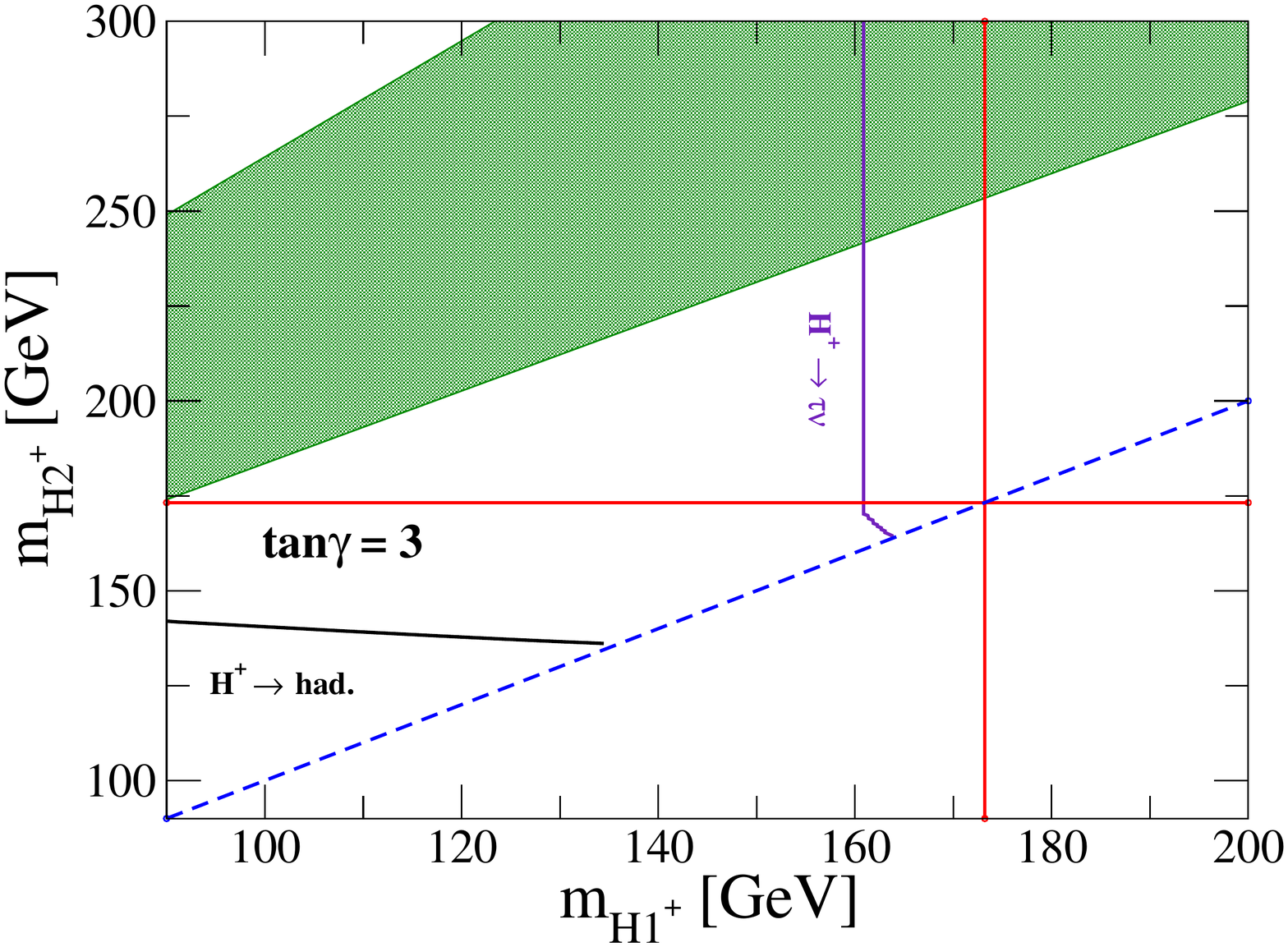}\hspace{-5mm}
\includegraphics[width=50mm]{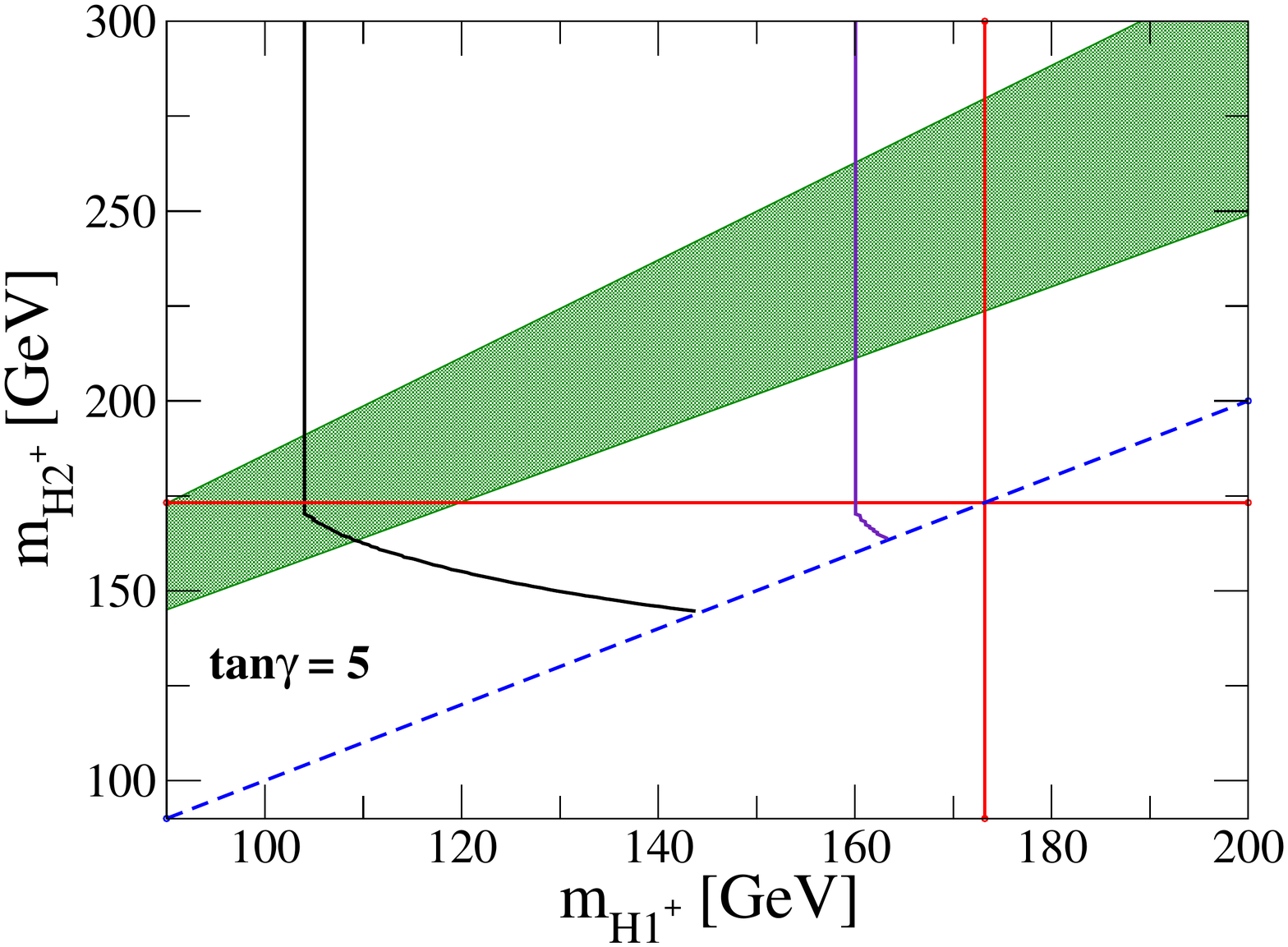}\hspace{-5mm}
\includegraphics[width=50mm]{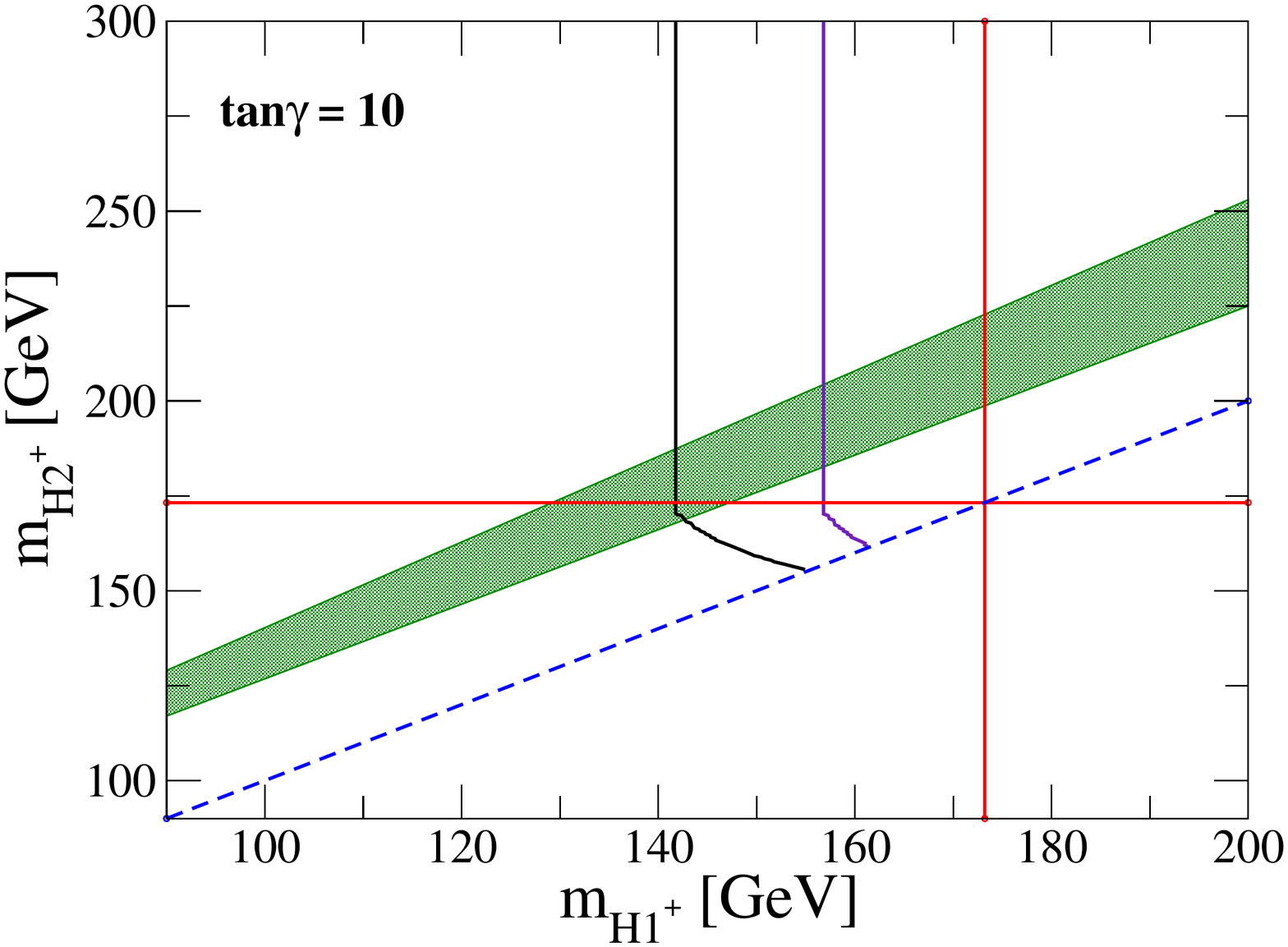} 
\caption{Parameter regions allowed by the $B\to X_s\gamma$ data and the direct searches at the LHC on the $m_{H_1^\pm}$--$m_{H_2^\pm}$ plane in the Type-Y (upper panels)
and the Type-Z (lower panels) 3HDM in the case of $\tan\beta=2$ and $\theta_C=-\pi/4$. 
We take $\tan\gamma$ to be 3, 5 and 10 from the upper left to lower right panels. 
Green shaded regions are allowed by the $B\to X_s\gamma$ data at 95\% CL. 
The right region from the black and purple curve is allowed by 
the search for $H^\pm \to \tau^\pm\nu$ and $H^\pm \to jj$ at the LHC. 
}
\label{const1}
\end{center}
\end{figure}

In contrast, in 3HDMs, the situation can be quite different from that in 2HDMs, namely,  
even in the Type-II like structure of the Yukawa interaction, we can take the combination of $\sum_a X_aY_a$ to be negative
due to the non-zero mixing of $H_1^\pm$ and $H_2^\pm$.  
We note that the predictions in the Type-X (Type-Y and Type-Z) Yukawa interaction are the same as that in the Type-I (Type-II). 
The other flavour constraints such as $Z\to b\bar{b}$, $B\to \tau\nu$ and $\tau\to \mu\nu\nu$ have been discussed in Ref.~\cite{Logan}
assuming that $H_2^\pm$ are much heavier than $H_1^\pm$. These typically do not give severe constraints when $\tan\beta\gtrsim 1$ is taken.

Next, we take into account the constraints from direct searches for charged Higgs bosons at the LHC. 
In Refs.~\cite{taunu} and \cite{jj}, the search for charged Higgs bosons decaying into $\tau\nu$ and $cs$ 
has been performed using the LHC Run-I data, respectively. 
From no significant excess of the event rates expected in the SM, the upper limits on BR($t\to H^\pm b$)$\times$BR($H^\pm \to \tau^\pm \nu$)
and BR($t\to H^\pm b$)$\times$BR($H^\pm \to jj$) have been presented for $m_{H^\pm} < m_t$.

In Fig.~\ref{const1}, we show the allowed parameter space by $B\to X_s \gamma$ and the direct searches at the LHC
on the $m_{H_1^\pm}$ and $m_{H_2^\pm}$ plane in the Type-Y (upper panels) and Type-Z (lower panels) 3HDM. 
It is clearly seen that the left-bottom area: $m_{H_{1,2}^\pm} < m_t$ is not allowed by $B\to X_s \gamma$ or the direct search, while 
there are allowed regions on the left-up area: $m_{H_1^\pm} < m_t$ and $m_{H_2^\pm}> m_t$. 
We note that when $m_{H_1^\pm}<m_t$, the BR of $H_1^\pm \to cb$ is given to be 
about 43\%, 68\% and 82\% (4\%, 16\% and 50\%) for the case with $\tan\gamma=3,~5$ and 10, respectively, in the Type-Y (Type-Z) 3HDM. 
Therefore, we find that the light charged Higgs boson scenario with dominant decay channel $H_1^\pm \to cb$ to be possible in the 3HDM, and 
this can be the smoking gun signature to identify these 3HDMs. 

\section{Conclusions and Discussions}

We have discussed the properties of charged Higgs bosons in 2HDMs and 3HDMs in the scenario based on NFC. 
We have shown that in the 3HDMs one of the charged Higgs bosons can be lighter than the top mass without contradiction with the $B\to X_s\gamma$ data 
and the direct search for charged Higgs bosons at the LHC Run-I experiment. 
It has been clarified that the decay mode $H_1^\pm \to cb$ can be dominant in the allowed parameter region in the Type-Y and Type-Z 3HDMs, which 
cannot be realized in the 2HDMs because of the constraint from $B\to X_s\gamma$. 
Therefore, the process, e.g., $pp \to t\bar{t} \to H^+ b W^- \bar{b}\to bb\bar{b}\ell^-\nu$ can be the smoking gun signature to identify the 3HDM. 
In this process, we can use the third $b$-jet tagging to reduce the background, which cannot be applied in the $H^\pm \to cs$ mode. 
Using this $b$-jet tagging, 
the signal significance is expected to be improved by about factor 2 as estimated in Ref.~\cite{cb}. 
In fact, recently the search for charged Higgs bosons decaying into $H^\pm \to cb$ has been performed at the LHC using the third $b$-tagging~\cite{cb_exp}, 
and it has been shown that 
the stronger limit on BR($t\to H^\pm b)\times$BR($H^\pm \to cb$) is taken as compared to that from the $H^\pm \to jj$ search without the third $b$-tagging. 

Finally, apart from the phenomenology of the charged Higgs bosons, 
let us briefly comment on the couplings of the SM-like Higgs boson ($h$) in the 3HDM. 
In Ref.~\cite{Yagyu}, it has been shown that the pattern of the deviation in the Yukawa couplings of $h$ 
can be completely different from that in the 2HDM within the framework of NFC. 
One of the remarkable points is that in the 3HDM, 
it is possible to obtain a prediction of the correlation between the deviation in the $h\tau^+\tau^-$ and $hb\bar{b}$ couplings which cannot be realized in the 2HDM even if 
we take into account the one-loop corrections to the Yukawa coupling~\cite{THDM-loop}. 
Therefore, we may also be able to find an indirect evidence for 3HDMs from the precise measurements of the Higgs boson couplings.

\end{document}